# Wafer-scale Epitaxial Graphene Growth on the Si-face of Hexagonal SiC (0001) for High Frequency Transistors

Christos Dimitrakopoulos,[a] Yu-Ming Lin, Alfred Grill, Damon B. Farmer, Marcus Freitag, Yanning Sun, Shu-Jen Han, Zhihong Chen, Keith A. Jenkins, Yu Zhu, Zihong Liu, Timothy J. McArdle, John A. Ott, Robert Wisnieff, and Phaedon Avouris

*IBM T. J. Watson Research Center, Yorktown Heights, NY 10598, USA*

**Abstract -** Up to two layers of epitaxial graphene have been grown on the Si-face of two-inch SiC wafers exhibiting room-temperature Hall mobilities up to 1800 $cm^2\ V^{-1}\ s^{-1}$, measured from ungated, large, 160 μm x 200 μm Hall bars, and up to 4000 $cm^2\ V^{-1}\ s^{-1}$, from top-gated, small, 1 μm x 1.5 μm Hall bars. The growth process involved a combination of a cleaning step of the SiC in a Si-containing gas, followed by an annealing step in Argon for epitaxial graphene formation. The structure and morphology of this graphene has been characterized using AFM, HRTEM, and Raman spectroscopy. Furthermore, top-gated radio frequency field effect transistors (RF-FETs) with a peak cutoff frequency $f_T$ of 100 GHz for a gate length of 240 nm were fabricated using epitaxial graphene grown on the Si face of SiC that exhibited Hall mobilities up to 1450 $cm^2\ V^{-1}\ s^{-1}$ from ungated Hall bars and 1575 $cm^2\ V^{-1}\ s^{-1}$ from top-gated ones. This is by far the highest cut-off frequency measured from any kind of graphene.

[a] Electronic mail: dimitrak@us.ibm.com

## I. INTRODUCTION

Graphene is an atomically thin layer comprising carbon atoms arranged in a two-dimensional (2D) hexagonal lattice. It is a 2D semiconductor that exhibits a linear relationship between electronic energy and 2D momentum, and the carriers in it are modeled as relativistic Dirac Fermions.[1,2] Graphene exhibits remarkable transport properties[3,4,5] including carrier mobilities of the order of 200,000 cm$^2$ V$^{-1}$ s$^{-1}$ at cryogenic temperatures in suspended graphene.[6] As a result, it is a good candidate for future high-frequency field effect transistor (FET) applications. Although a macroscopic, 2D graphene sheet is generally a zero-band-gap semiconductor, it has potential to be used as the active channel in FETs for analog, radio frequency (RF) applications such as low-noise amplifiers (LNA).

The 2D nature of graphene facilitates the fabrication of planar graphene devices and integrated circuits using standard semiconductor industry processes, a major advantage over one-dimensional structures such as carbon nanotubes (CNT). Graphene has been produced by exfoliation of graphite,[7] epitaxially on SiC by high temperature decomposition of the latter,[8] or on metals.[9,10,11]

Although high quality graphene flakes can be obtained by repetitive exfoliation of graphite, this is not a process that can be used in large scale fabrication, mainly due to the fact that no suitable method has been demonstrated yet that enables the formation of structurally coherent graphene over wafer-scale areas, or even the formation of large arrays of small-size graphene flakes arranged with a unique and predictable azimuthal orientation on a substrate. Growing graphene by subtractive epitaxy on semi-insulating SiC substrates, offers the advantage of structural coherence over large areas, because the azimuthal orientation of graphene is governed, to a large degree, by the crystal structure of the substrate surface. This is especially true in the case of the Si-face surface of SiC.[12, 13]

In this paper, we report the growth of one to two layers of epitaxial graphene on the Si-face of two-inch SiC wafers, its structure and morphology, and its room-temperature Hall mobility measured on large ungated Hall bars and small gated Hall bars. Furthermore, we provide for the first time a detailed account on the graphene growth process and characterization, device fabrication, and device characteristics for the recently reported record-breaking radio frequency field effect transistors (RF-FETs) comprising such epitaxial graphene on SiC, and exhibiting record cutoff frequency $f_T$ up to 100 GHz for a gate length of 240 nm.[14]

## II. EXPERIMENTAL PROCEDURES

We grew graphene on the Si face of either high purity semi-insulating (HPSI) 4H(0001) 2" SiC wafers or semi-insulating (SI) 6H(0001) 2" SiC wafers. The wafers of both polytypes had a chemical mechanical polished (CMP) epitaxy-ready surface on their Si polar face.

Graphene was grown in a UHV chamber (base pressure of ~ $3 \times 10^{-10}$ Torr) equipped with a custom-designed, inductively heated graphite susceptor hot-zone that can accommodate two-inch wafers. The shape, dimensions, shielding, and materials of the hot zone were optimized to produce uniform heating of two-inch wafers up to the required process temperatures. The temperature was measured with a two-color pyrometer, which is connected in a feedback loop comprising a programmable temperature controller and a 12 KW RF power supply.

For the samples that we used for the RF-FET devices, we grew graphene on 4H(0001) SiC wafers, by combining for the first time a cleaning step under a Si-containing gas followed by annealing/graphenization at a higher temperature in Argon. The SiC wafer was cleaned mainly from oxide contamination by annealing at 810 °C under disilane flow (20% disilane in He). Chemical vapor cleaning processes involving Si-containing gas molecules have been shown in the past to effectively remove oxides from the surface of SiC[15, 16] mainly by converting $SiO_2$ to SiO, the latter being a more volatile oxide. After the cleaning step, the SiC wafer was annealed at 1450 °C for 2 min under Ar flow at a pressure of $3.5 \times 10^{-4}$ Torr and then was allowed to cool down in Ar.

The sample that showed the highest gated Hall mobility in Table III, was grown on the Si-face of a semi-insulating 6H(0001) SiC wafer. The process sequence used for this sample comprised an extra annealing step (compared to the original process described above) in 20% disilane in He at T=1140 °C at pressure $P=8 \times 10^{-7}$ Torr, i.e under conditions that favor the $\sqrt{3} \times \sqrt{3}$ surface reconstruction of SiC.[16] After the cleaning step at 810 °C and the 1140 °C anneal (both under disilane flow), the SiC wafer was annealed at 1450 °C for 10 min under Ar flow at a pressure of $3.5 \times 10^{-4}$ Torr and then was allowed to cool down in Ar.

Ungated Hall bar devices (200 µm x 160 µm) were fabricated by depositing Ti/Pd/Au contacts on top of blanket graphene using optical lithography and lift-off, followed by graphene patterning to the appropriate Hall bar shape using another optical lithography step, oxygen RIE, and wet striping of the resist. For the smaller Hall devices (1.5 µm x 1.0 µm) e-beam lithography was used instead of optical lithography.

Atomic layer deposition (ALD) was used for the deposition of the high-k gate dielectric used in the fabrication of top-gated graphene Hall bars and RF-FETs. In order to enable uniform oxide coverage, a 10-nm-thick layer of polyhydroxystyrene-based polymer was spun first on the graphene surface to act as a nucleation layer, followed by ALD of a 10-nm-thick layer of $HfO_2$.[17] Finally, a Pd/Au (20 nm/40 nm) top-gate electrode was deposited and patterned by lift-off.

The samples were characterized by micro-Raman spectroscopy, tapping mode atomic force microscopy (AFM) and high resolution transmission electron microscopy (HRTEM). Micro-Raman spectroscopy was performed at room temperature with a Triax 322 spectrometer (Horiba

Jobin Yvon) at 514.5 nm, using a 1200 line, 500nm blaze grating. The light was sent through a 100x objective with 0.8 NA to form a ~500 nm focal spot in the middle of the Hall bars. Measurements were performed with sub-mW incident power to avoid heating. A SiC background, recorded on the same wafer in an area where the graphene had been removed, was subtracted from the raw spectrum, leaving behind only peaks due to the graphene overgrowth.

TEM samples were prepared using a dual-beam focused ion beam (FIB) in-situ lift-out technique using 30 keV Ga ions. Prior to ion milling, a sacrificial film was deposited to preserve the ion damage from the focused ion beam. The samples were cleaned by lower energy (a few keV) Ga ions. Bright-field HRTEM images of the interface between graphene and SiC were taken using a JEOL-3000F TEM with an accelerating voltage of 300 kV.

## III. RESULTS AND DISCUSSION

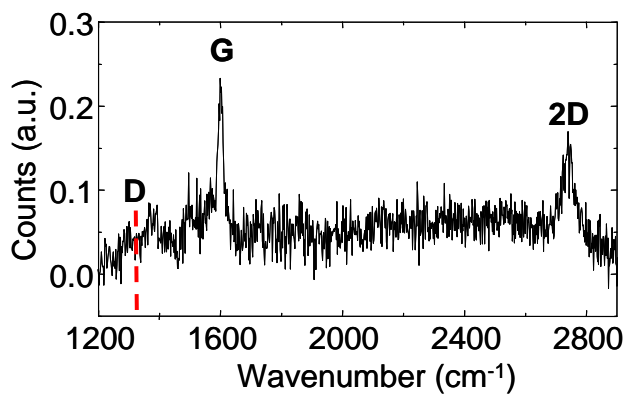

**Figure 1:** Raman spectrum from a (HPSI) 4H(0001) SiC wafer after graphene formation (SiC background was subtracted).

Figure 1 shows the Raman spectrum of a graphene on 4H(0001) SiC sample prepared as described above, after subtraction of the SiC background. The main features observed in this spectrum are the peaks labeled G and 2D. We also label the expected position of the D peak (at about 1325 cm$^{-1}$, marked with red dashed line), which is attributed to defects in graphene (e.g. edges or disordered regions).[18] This band is forbidden in defect-free graphene and graphite, and its absence is an indication of high quality graphene and minimal contribution from domain edges. The G peak centered at 1600 cm$^{-1}$ is a first order band due to the doubly degenerate zone center E$_{2g}$ optical phonon.[19] The 2D-phonon band at 2739 cm$^{-1}$, is a second order peak due to two zone boundary optical phonons. The relative intensities of the G and 2D peaks and the weak absorption of the SiC Raman signal (not shown here) due to the graphene overlayer indicates 1-2 layer graphene.

Figure 2a shows a tapping mode AFM height image taken ex situ from this sample. The surface morphology is uniform over the whole wafer. The wafer surface is vicinal but pits appear on the terraces. Nevertheless, long continuous strips of unpitted material remain on each of the terraces. Figure 2b shows the corresponding AFM phase image. AFM phase image contrast has been shown to be created between the $\sqrt{3} \times \sqrt{3}$ surface reconstruction of SiC and the buffer layer (the $6\sqrt{3} \times 6\sqrt{3} \cdot R30$ C-rich reconstruction of hexagonal SiC), as well as between the buffer layer and graphene.[16] However, we have obtained evidence that AFM phase image contrast is also created when different phases exist immediately below the top-most graphene layer (e.g. graphene on buffer versus graphene on graphene) due to different rates of energy dissipation between the tip and the sample for the different layer stacks (phase contrast arises from differences in the energy dissipation between the tip and the sample[20]). The brown regions in Fig. 2b are mostly linked to surface pits created by SiC decomposition, thus are expected to have more graphene than the yellow regions. If the yellow regions corresponded to buffer layer and brown regions to graphene regions, there would be no electrical conduction, as the brown regions are discontinuous and the buffer layer is not conductive. Thus, based on the excellent sheet conductivity and mobility properties measured from this sample (see Table I below) the yellow regions should correspond to 1 monolayer (ML) of graphene and the brown regions to 2 ML of graphene. Based on this, we can conclude that one continuous graphene layer has been grown conformally over steps and pits on the SiC surface of this sample, and in addition, a second, discontinuous graphene layer has been grown (brown regions) under the top graphene layer.

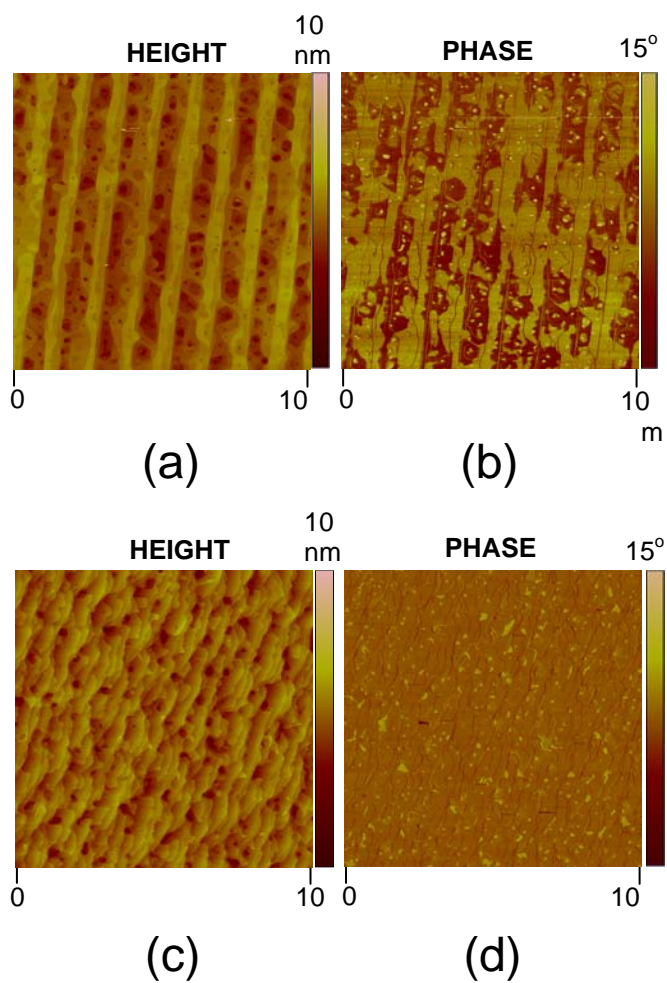

**Figure 2**: (a) and (b): Tapping mode AFM images from 4H(0001) SiC wafer after graphene formation using the original growth process. (a) Height image; (b) Phase image from the same area of graphene as in (a): Yellow vs. brown regions: 1 monolayer (ML) graphene vs. 2 ML

graphene (see discussion in text). One continuous graphene layer grown conformally over steps and pits. (c) and (d): Tapping mode AFM images from 6H(0001) SiC wafer after graphene formation using the modified growth process. (c) Height image; (d) Phase image from the same area of graphene as in (c): Yellow vs. brown regions are observed that as in (b) they should correspond to 1 ML graphene vs. 2 ML, respectively, however in this sample, the brown region covers the majority of the surface.

Figures 3a and b show HRTEM cross sections taken from the ungated Hall bar device area of the same sample. The position of the graphene layers is indicated by the dark band sandwiched between the two bright white bands, due to the slightly underfocused conditions used.[21] Up to a maximum of two layers of graphene (one plus a second possibly incomplete layer) appear to have been grown over SiC. Interestingly, the graphene layers seem to have been grown conformally over a two-bilayer step (~0.5 nm high) of the SiC vicinal surface (Fig. 3b). This is a direct confirmation of previous evidence provided by scanning tunneling microscopy (STM) that graphene layers are continuous over SiC steps.[8,22] It also explains why the existence of step edges had a minor effect on the mobility of multilayer graphene FETs having their current flowing parallel or perpendicular to the steps.[23]

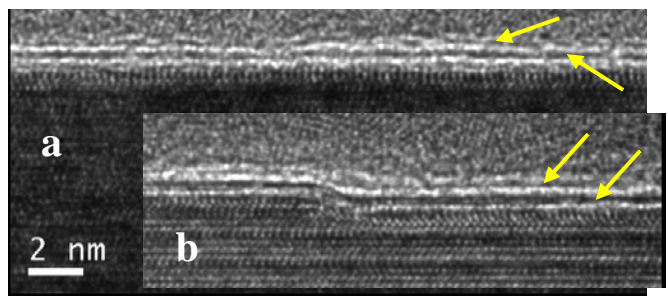

**Figure 3**: HRTEM cross sections a) Graphene is the dark stripe between the thick bright stripes, which are attributed to the spacing between graphene and adjacent layers in this focus condition.

One continuous graphene layer, and possibly a second graphene layer can be observed (arrows).; b) Graphene layer conformally covering a two-bilayer step on the SiC surface.

Standard, ungated Hall bar devices were used to measure the Hall mobility of graphene. The basic assumptions of the standard Hall measurement method (one kind of dominant charge carrier, the sheet resistance parallel to the constant current flow being independent of applied magnetic field) are valid for one to two layers of epitaxial graphene grown on the Si-face of SiC. Sheet resistance, $R_s$, was calculated from Hall bar devices built on as-prepared graphene using the equation:

$$R_s = \frac{(V_{X_2} - V_{X_1})w}{Id} \tag{1}$$

where $w$ is the Hall bar channel width (200 µm), d is the distance between voltage leads $X_2$ and $X_1$ (160 µm), and $I$ is a known current flowing between leads 1 and 2 (Fig. 4 shows a micrograph of the Hall bar device used for these measurements).

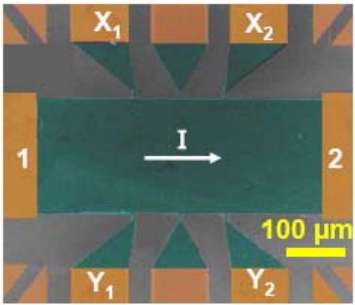

**Figure 4**: Image of photolithographically patterned graphene Hall bar.

The 2D carrier density ($n_s$) is derived by the equation:

$$n_s = -\frac{1}{eR_H} \tag{2}$$

where $e$ is the electron charge, $R_H$ is the Hall coefficient given by

$$R_H = \frac{V_{Y_1} - V_{X_1}}{IB} \qquad (3)$$

and $B$ is the magnetic field applied perpendicular to the Hall bar plane.[24] The Hall mobility ($\mu_H$) is calculated by

$$\mu = \frac{1}{n_s e R_s}. \qquad (4)$$

Measurements from ungated Hall bar devices were performed on various devices built over one quarter of the same two-inch wafer used in the characterization above. Mobilities varied from 1070 to 1450 cm$^2$ V$^{-1}$ s$^{-1}$ with carrier densities between $3.6 \times 10^{12}$ and $2.8 \times 10^{12}$ cm$^{-2}$ respectively. The majority carriers were electrons as shown by the positive slope of the plot of the voltage difference between electrodes X2 and Y2 of the Hall bar (indicated by $V_{xy}$) vs. the magnetic field $B$, in Figure 5, that corresponds to Hall bar (4,1) whose electrical properties are described in Table I. A perfectly linear plot corroborates the validity of the assumptions employed in the standard Hall mobility measurement for one to two layer graphene grown on the Si-face of SiC, which were described above.

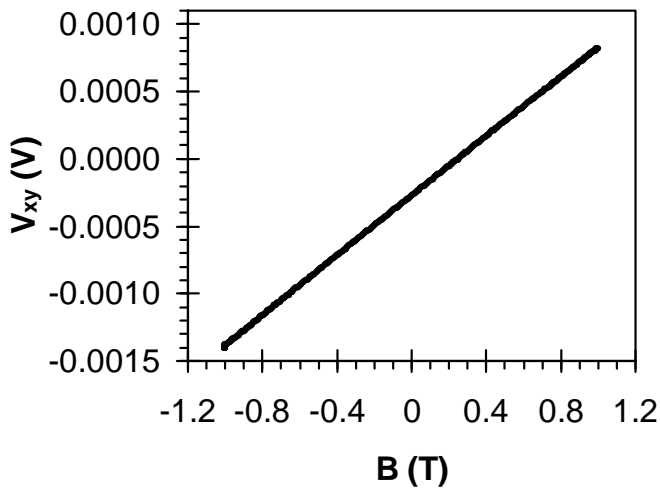

**Figure 5**: Plot of the voltage difference $V_{xy}$ between electrodes X2 and Y2 of the Hall bar vs. the magnetic field B. A perfectly linear plot proves the validity of the assumptions used in the standard Hall mobility measurement method for one to two layer graphene grown on the Si-face of SiC (e.g. dominance of a single type of majority carriers, and constant $R_s$ with varying magnetic field).

Table I shows the properties of three such devices. It should be stressed that the size of the Hall bar devices (200 µm x 160 µm) is large and that it contains approximately one hundred or more vicinal terraces of SiC. This means that either a single graphene domain larger than these dimensions occupies the entire Hall bar area, or that graphene domain boundaries, if existing, do not substantially affect the effective mobility of a multidomain graphene device, which would explain the high mobility measured.

**Table I:** Hall bar measurement results from as-prepared graphene on 4H(0001) SiC.

| Device (row, column) | 4,1 | 5,1 | 1,3 |
|---|---|---|---|
| Sheet Resistance (ohm/sq) | 1541 | 1624 | 1550 |
| Carrier Density (cm$^{-2}$) | 2.8x10$^{12}$ | 3.6 x10$^{12}$ | 3.0 x10$^{12}$ |
| Hall Mobility (cm$^2$ V$^{-1}$ s$^{-1}$) | 1450 | 1070 | 1330 |

Top-gated RF FETs were fabricated on the same SiC/graphene wafer to study the high frequency response of epitaxial graphene transistors. Adjacent to the RF FETs, top-gated Hall bar devices (inset of Figure 6) were built to study the effect of gate-modulated charge carrier concentration on Hall mobility. The Hall mobility was measured at discrete gate voltage $V_G$ values, as described above. Figure 6 shows Hall mobility vs. gate voltage $V_G$ for two devices. A maximum Hall mobility of 1575 cm$^2$ V$^{-1}$ s$^{-1}$ was measured at $V_G$= -4.5 V.

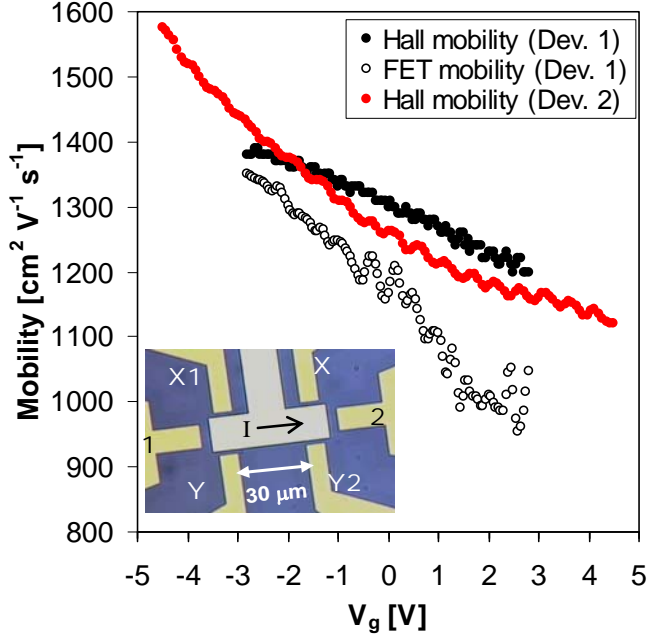

**Figure 6:** Hall and field effect mobility vs. gate voltage measured from two top-gated Hall bar devices.

Field-effect mobility, $\mu_{eff}$, was measured from the top-gated Hall bar device using the relation:

$$\mu_{eff} = \frac{\Delta G_s}{e \cdot \Delta n_s} \qquad (5)$$

where $\Delta G_s$ is the change in the sheet conductance, $\Delta n_s$ is the 2D carrier density modulation due to the change in $V_G$, because

$$\Delta G_s = \frac{\Delta I_{ds} \cdot L}{V_{ds} \cdot W} \qquad (6)$$

and

$$\Delta n_s \cdot e = C_i \cdot \Delta V_g . \qquad (7)$$

The capacitance per unit area $C_i$ was measured to be $1.3 \times 10^{-7}$ F cm$^{-2}$. It is important to note that using the 4-probe setup of the Hall bar device to measure $G_s = 1/R_s$, we avoided making any assumptions about the contact resistance of the FET, which would have introduced errors in the estimation of effective mobility. The distance between the voltage leads $X_2$ and $X_1$ of the top-gated Hall bar was $L=30$ μm, while the graphene bar width was $W=4$ μm. Based on the variation of the measured sheet conductance and the 2D charge density as a function of $V_g$ and using the equations shown above, the field-effect mobility vs. $V_g$ plot was derived and is shown on Figure 6 (device 1, open circles). Good agreement between Hall and field-effect mobilities is observed (within 10%). The highest field-effect mobility measured in this way was 1400 cm$^2$ V$^{-1}$ s$^{-1}$. The agreement between our Hall mobilities measured before and after the deposition of the gate dielectric stack, and the field effect mobilities is quite remarkable and unusual based on previous literature results.

In order to study the high frequency response of epitaxial graphene top-gated RF FETs we used the methodology described by Lin et al. [25] The high-frequency performance of a transistor for small-signal response is measured by S-parameter measurements up to 30 GHz, and it is mainly determined by the transconductance $g_m=dI_d/dV_g$.[25]

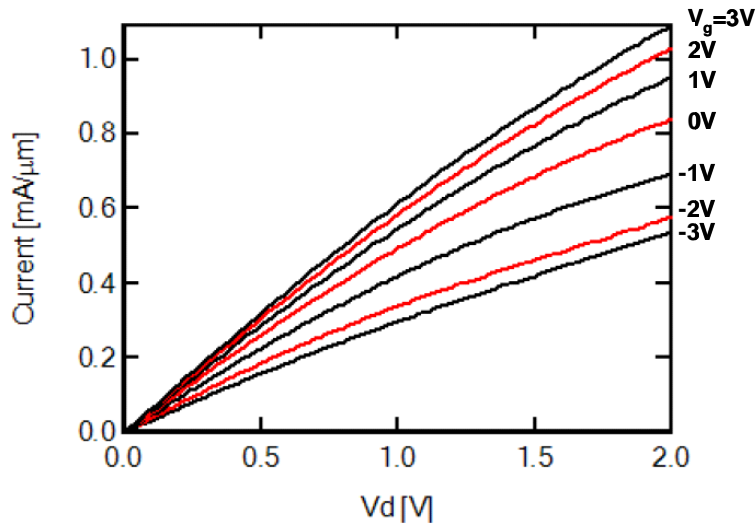

Figure 7: FET output characteristics for a device with $L_G=550$ nm.

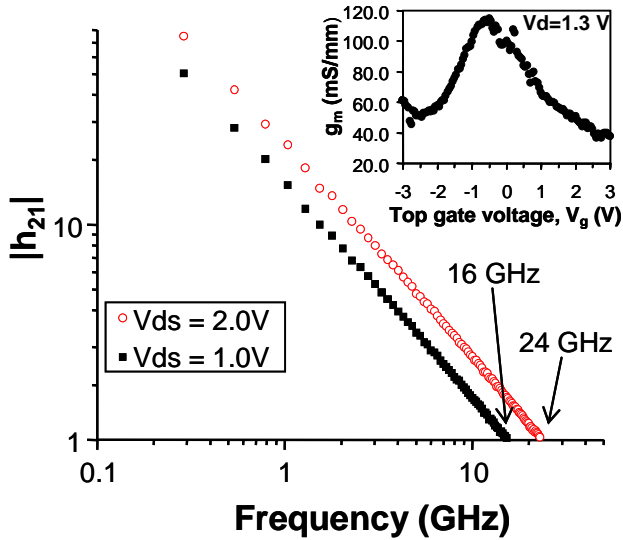

**Figure 8:** De-embedded current gain |h$_{21}$| vs. frequency for an RF FET with L$_G$=550 nm. A cut off frequency f$_T$ is determined to be 16 and 24 GHz for V$_d$ of 1.0 and 2.0 V, respectively. Inset: Transconductance normalized to channel width as a function of top gate voltage at V$_d$ =1.3 V for the same RF FET.

Figure 7 shows the FET output characteristics for a device with L$_G$=550 nm. The maximum on current measured from this n-type epitaxial graphene FET was 1.1 mA/μm at V$_g$=3 V and V$_d$=2 V. The inset of figure 8 shows a plot of transconductance g$_m$ normalized to channel width as a function of top gate voltage at V$_d$=1.3 V for a RF FET with gate length L$_G$ = 550 nm and channel width W=30 μm. The magnitude of g$_m$ is strongly dependent on V$_G$, but its sign is positive throughout the V$_g$ range used, showing that transport is dominated by electrons, in agreement with Hall measurements. For V$_d$ = 1.3 V, the peak g$_m$ is measured to be 115 mS/mm at V$_g$ = -0.5 V. Figure 8, which corresponds to one of the first devices fabricated and tested on this graphene wafer, shows the current gain h$_{21}$ as a function of frequency, after de-embedding the pad parasitics using specific short and open structures. The current gain exhibits the 1/f frequency dependence expected for a well-behaved FET. The cutoff frequency, defined as the frequency of unity current gain, is measured to be 16 and 24 GHz [31] for V$_d$ of 1.0 and 2.0 V, respectively, for the 550-nm-gate-length graphene FET. The highest $f_T$ we measured for devices built on the same wafer and having the same gate length $L_G$=550 nm was 53 GHz, as shown in figure 9. The variation in $f_T$ for several devices built on the same wafer and having the same gate length of 550 nm was between 20 and 53 GHz.

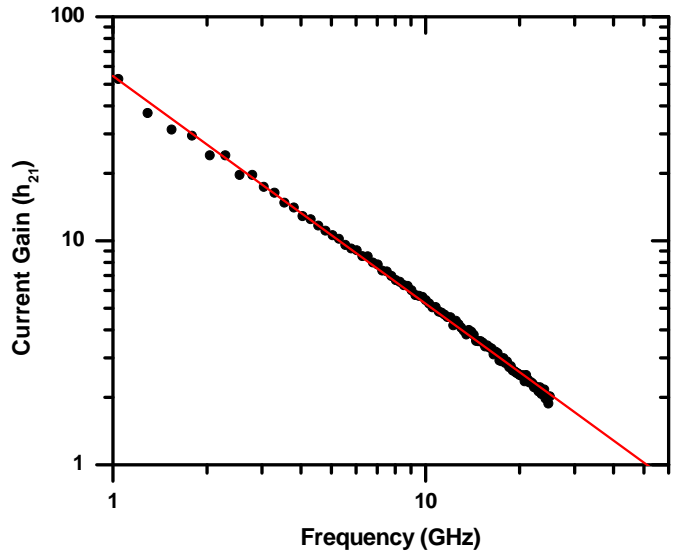

Figure 9: De-embedded current gain $|h_{21}|$ vs. frequency for a RF FET with $L_G$=550 nm. Exhibiting the maximum cut off frequency $f_T$ = 53 GHz that we measured from devices of this size.

As expected from previous results on exfoliated graphene FETs that showed that $f_T$ is proportional to $(1/L_G)^2$,[25] FETs built on the same epitaxial graphene on SiC wafer and having a shorter gate length, $L_G$=240 nm, produced higher peak cutoff frequencies as shown in Figure 10 (some of the data in Figure 10 were first reported in reference 14 recently). The maximum $f_T$ we measured from these devices was 100 GHz.[14] This is the highest cutoff frequency reported from any kind of graphene based FETs. The highest $f_T$ value reported previously for epitaxial graphene FETs on SiC was 4.4 GHz,[28] while the highest $f_T$ value reported previously for an exfoliated graphene flake based FETs was 50 GHz. [26]

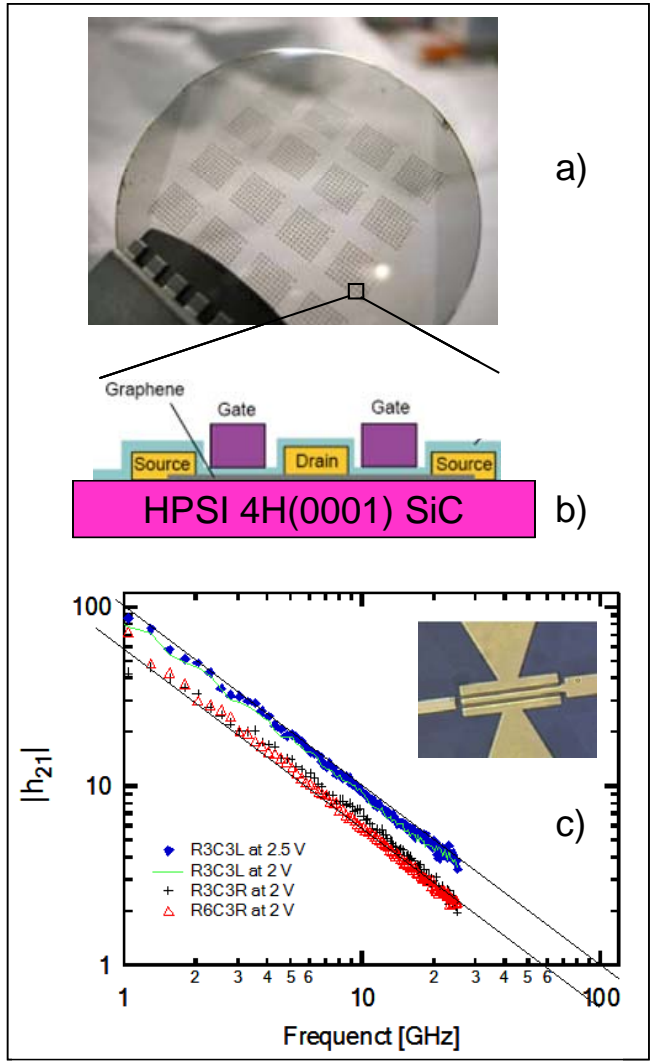

Figure 10: a) Devices fabricated on epitaxial graphene on the Si-face of a 2 inch HPSI SiC 4H(0001) wafer. b) Cross section of epitaxial graphene RF-FET device. c) De-embedded current gain $|h_{21}|$ vs. frequency for three RF FETs with $L_G$=240 nm exhibiting a range of cutoff frequencies $f_T$ between 60 and 100 GHz. All three devices are tested at drain biases of 2 V but the devices that showed the highest $f_T$ at 2 V was also tested at $V_D$ = 2.5 V resulting in $f_T$ = 100 GHz (Some of this data was reported in reference 14).

Since the fabrication and testing of these RF devices, we have improved the ungated and gated Hall mobility of epitaxial graphene on the Si-face of SiC wafers by optimizing the process and materials combination that we used (see experimental section). As a result we have measured Hall mobilities from 1760 to 1804 cm$^2$ V$^{-1}$ s$^{-1}$ from large, ungated 160 μm x 200 μm Hall bar devices fabricated with epitaxial graphene as shown on Table II. This epitaxial graphene was grown on the Si face of a semi-insulating 6H(0001) SiC and the process sequence comprised an extra annealing step in 20% disilane in He at T=1140 $^o$C. Then, the SiC wafer was annealed at 1450 $^o$C for 10 min under Ar flow at a pressure of 3.5x10$^{-4}$ Torr.

**Table II:** Hall bar measurement results from as-prepared graphene on semi-insulating 6H(0001) SiC.

| **Device (row, column)** | (2,5) | (3,4) | (2,3) |
|---|---|---|---|
| **Sheet Resistance (ohm/sq)** | 1451 | 1692 | 1715 |
| **Carrier Density (cm$^{-2}$)** | -2.4E12 | -2.1E12 | -2.1E12 |
| **Hall Mobility (cm$^2$ V$^{-1}$ s$^{-1}$)** | 1804 | 1765 | 1775 |

Figure 2c shows tapping mode AFM height image from epitaxial graphene grown on the Si-face of a semi-insulating 6H(0001) substrate with the aforementioned process. There is less pitting than that shown in Figure 2a, corresponding to graphene on 4H(0001) SiC grown without the extra annealing step in 20% disilane in He at T=1140 $^o$C. The phase image in Figure 2d is from the same area of graphene. Yellow vs. brown regions, like in Figure 2b, should correspond to 1 ML graphene vs. 2 ML, respectively. However in this sample, the brown region covers the majority of the surface. As a result of the better coverage with two graphene layers, compared to the sample of Figure 2a,b, the ungated Hall mobility has increased to 1800 cm$^2$ V$^{-1}$ s$^{-1}$ from 1450 cm$^2$ V$^{-1}$ s$^{-1}$.

Figure 11a shows the Raman spectrum of graphene on SiC 6H(0001) (same sample corresponding to Table II and Figure 2c,d) after SiC background subtraction. Applying the same analysis as for the spectrum in Figure 1, we can conclude from the relative intensities of the G and 2D peaks and the weak absorption of the SiC Raman signal (not shown here) due to the graphene overlayer that the graphene coverage is about 1-2 layers (but less than three). We took Raman spectra in three positions from the center to the edge of the wafer and plotted the intensity of the 2D peak (characteristic of graphene) for the three positions (Figure 11b). Very good thickness uniformity center to edge can be deduced.

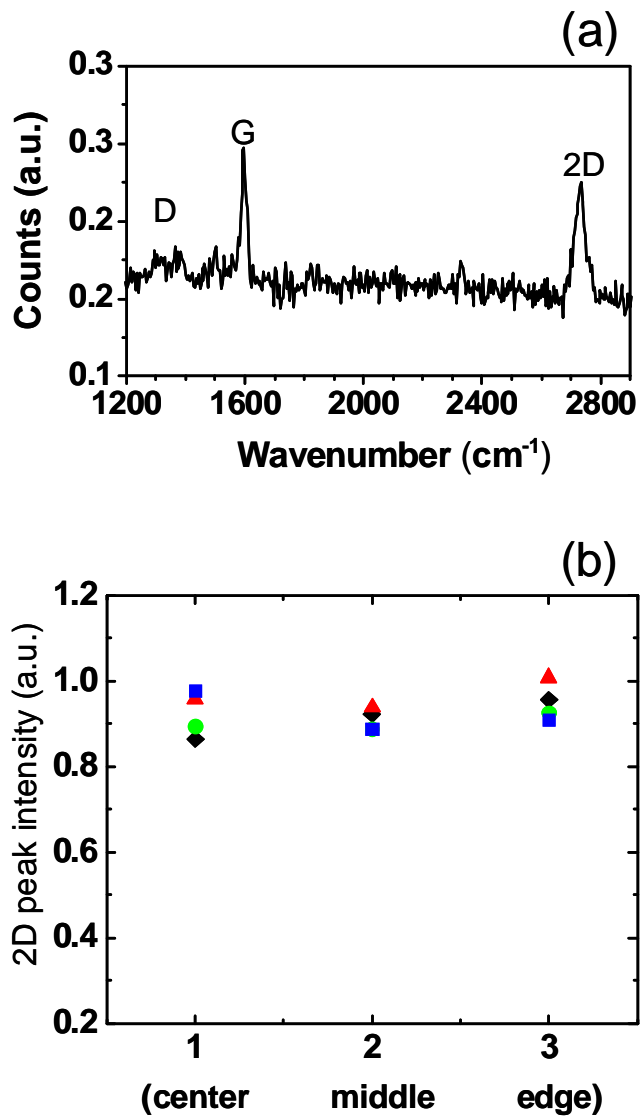

Figure 11: (a) Raman spectrum of graphene on SiC 6H(0001) (same sample corresponding to Table II and Figure 2c,d) after SiC background subtraction. (b) Plot of the 2D peak intensity of the Raman spectrum of the same sample for three different positions on a quarter of a 2-inch wafer (close to the center, intermediate position between center and edge, and close to the edge). For each position, several spectra separated by about 10 μm from each other were taken and plotted.

To test the effect of pits on the measured Hall mobility at room temperature, we fabricated small, gated Hall bars (size 1 μm x 1.5 μm) in a targeted, pit-free area of graphene on SiC (same sample corresponding to Table II). Figure 12 shows the room-temperature Hall mobility measured from such a Hall bar as a function of top-gate voltage. The mobility increases as the concentration of electrons in graphene is reduced, up to a value $\mu_{Hall}$ just above 4000 cm$^2$/Vs at $V_{TG}$=-7.2 volts, where the electron density is calculated to be just above $1.0 \times 10^{11}$ cm$^{-2}$ (red symbols). At more negative top-gate voltages (lower electron concentrations) and as the Dirac point is approached, ambipolar transport starts becoming important and the equations used to model the device behavior are not expected to be valid (grey symbols). The mobility measured at electron density of $2.27 \times 10^{12}$ cm$^{-2}$ (similar to the values reported in Table II from large Hall bars), is 1750 cm$^2$/Vs. This value is similar to the Hall mobility measured from the large Hall bars (Table II), the latter being an average (effective) mobility for graphene on flat and pitted areas, and hundreds of vicinal steps and terraces. This proves that pits on the SiC surface do not have an important effect on the Hall mobility of the graphene grown over them.

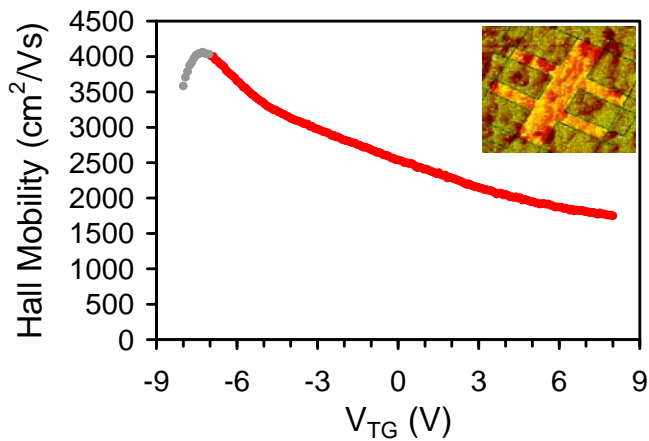

Figure 12: Plot of room-temperature Hall mobility vs. top-gate voltage.

Table III lists the room temperature Hall mobilities (published [27, 28, 29, 30, 31, 32] and from this work) from graphene grown on the Si face of hexagonal SiC. The ungated Hall mobility we report is higher than all but one previously published result, while the gated Hall mobility reported here is by far the highest Hall mobility (gated or ungated bars) reported from graphene grown on the Si face of SiC.

**Table III:** Room temperature Hall mobilities (published and from this work) from graphene grown on the Si face of hexagonal SiC

| Hall mobility ($cm^2$ $V^{-1}$ $s^{-1}$) | Reference |
|---|---|
| 1800 [b] <br> 4000 [c, d] | This work |
| 2400 | 32 |
| 1640 | 33 |

[b] Hall bar size 160 μm x 200 μm.
[c] Hall bar size 1μm x 1.5 μm.

| | |
|---|---|
| 1600 [d] | 30 |
| ~ 1600 | 34 |
| 1450 [b]  1575 [d] | 31 and this work |
| 1150 | 35 |
| 1000 | 28 |
| 900 | 27 |
| 860 | 29 |

[d] From gated Hall bar

## IV. CONCLUSION

In conclusion, we have grown up to two layer epitaxial graphene on the Si-face of two inch diameter semi-insulating 4H(0001) SiC wafers, and fabricated and tested ungated and gated Hall bar devices as well as RF FETs using such graphene as the active layer. Room-temperature Hall mobilities up to 1800 cm$^2$ V$^{-1}$ s$^{-1}$, measured from ungated, large, 160 μm x 200 μm Hall bars, and up to 4000 cm$^2$ V$^{-1}$ s$^{-1}$, from top-gated, small, 1 μm x 1.5 μm Hall bars. Record peak cutoff frequency, $f_T$, up to 100 GHz was measured from top-gated epitaxial graphene FETs with gate length $L_G$=240 nm, which is the highest cutoff frequency reported from FETs with any kind of graphene channels. The graphene used for the latter RF-FETs, had room-temperature Hall mobilities up to 1450 cm$^2$ V$^{-1}$ s$^{-1}$. Thus, based on the higher mobilities demonstrated by our most recent epitaxial graphene samples grown with our modified graphene growth process, there is considerable potential for achieving even higher RF-FET performance in the near future.

**ACKNOWLEDGMENTS**

This work was supported by DARPA under Contract FA8650-08-C-7838 (CERA program).

The authors are grateful to Dr. Chun-Yung Sung for the guidance and administration of the CERA project at IBM.

The views, opinions, and/or findings contained in this article/presentation are those of the author/presenter and should not be interpreted as representing the official views or policies, either expressed or implied, of the Defense Advanced Research Projects Agency or the Department of Defense.

DARPA Distribution Statement A : Approved for Public Release, Distribution Unlimited.

**Table I:** Hall bar measurement results from as-prepared graphene on 4H(0001) SiC.

| Device (row, column) | 4,1 | 5,1 | 1,3 |
|---|---|---|---|
| Sheet Resistance (ohm/sq) | 1541 | 1624 | 1550 |
| Carrier Density (cm$^{-2}$) | 2.8x10$^{12}$ | 3.6 x10$^{12}$ | 3.0 x10$^{12}$ |
| Hall Mobility (cm$^2$ V$^{-1}$ s$^{-1}$) | 1450 | 1070 | 1330 |

**Table II:** Hall bar measurement results from as-prepared graphene on semi-insulating 6H(0001) SiC.

| Device (row, column) | (2,5) | (3,4) | (2,3) |
|---|---|---|---|
| Sheet Resistance (ohm/sq) | 1451 | 1692 | 1715 |
| Carrier Density (cm$^{-2}$) | -2.4E12 | -2.1E12 | -2.1E12 |
| Hall Mobility (cm$^2$ V$^{-1}$ s$^{-1}$) | 1804 | 1765 | 1775 |

**Table III:** Room temperature Hall mobilities (published and from this work) from graphene grown on the Si face of hexagonal SiC

| Hall mobility (cm$^2$ V$^{-1}$ s$^{-1}$) | Reference |
|---|---|
| 1800 [a]<br>4000 [b, c] | This work |
| 2400 | 32 |
| 1640 | 33 |
| 1600 [c] | 30 |
| ~ 1600 | 34 |
| 1450 [a]<br>1575 [c] | 31 and this work |
| 1150 | 35 |
| 1000 | 28 |
| 900 | 27 |
| 860 | 29 |

[a] Hall bar size 160 μm x 200 μm.
[b] Hall bar size 1μm x 1.5 μm.
[c] From gated Hall bar

**FIGURE CAPTIONS**

**Figure 1:** Raman spectrum from a (HPSI) 4H(0001) SiC wafer after graphene formation (SiC background was subtracted).

**Figure 2**: (a) and (b): Tapping mode AFM images from 4H(0001) SiC wafer after graphene formation using the original growth process. (a) Height image; (b) Phase image from the same area of graphene as in (a): Yellow vs. brown regions: 1 monolayer (ML) graphene vs. 2 ML graphene (see discussion in text). One continuous graphene layer grown conformally over steps and pits. (c) and (d): Tapping mode AFM images from 6H(0001) SiC wafer after graphene formation using the modified growth process. (c) Height image; (d) Phase image from the same area of graphene as in (c): Yellow vs. brown regions are observed that as in (b) they should correspond to 1 ML graphene vs. 2 ML, respectively, however in this sample, the brown region covers the majority of the surface.

**Figure 3**: HRTEM cross sections a) Graphene is the dark stripe between the thick bright stripes, which are attributed to the spacing between graphene and adjacent layers in this focus condition. One continuous graphene layer, and possibly a second graphene layer can be observed (arrows).; b) Graphene layer conformally covering a two-bilayer step on the SiC surface.

**Figure 4**: Image of photolithographically patterned graphene Hall bar.

**Figure 5**: Plot of the voltage difference $V_{xy}$ between electrodes X2 and Y2 of the Hall bar vs. the magnetic field B. A perfectly linear plot proves the validity of the assumptions used in the standard Hall mobility measurement method for one to two layer graphene grown on the Si-face of SiC (e.g. dominance of a single type of majority carriers, and constant $R_s$ with varying magnetic field).

**Figure 6:** Hall and field effect mobility vs. gate voltage measured from two top-gated Hall bar devices.

**Figure 7:** FET output characteristics for a device with $L_G$=550 nm.

**Figure 8:** De-embedded current gain $|h_{21}|$ vs. frequency for an RF FET with $L_G$=550 nm. A cut off frequency $f_T$ is determined to be 16 and 24 GHz for $V_d$ of 1.0 and 2.0 V, respectively. Inset: Transconductance normalized to channel width as a function of top gate voltage at $V_d$ =1.3 V for the same RF FET.

**Figure 9:** De-embedded current gain $|h_{21}|$ vs. frequency for a RF FET with $L_G$=550 nm. Exhibiting the maximum cut off frequency $f_T$ = 53 GHz that we measured from devices of this size.

**Figure 10:** a) Devices fabricated on epitaxial graphene on the Si-face of a 2 inch HPSI SiC 4H(0001) wafer. b) Cross section of epitaxial graphene RF-FET device. c) De-embedded current gain $|h_{21}|$ vs. frequency for three RF FETs with $L_G$=240 nm exhibiting a range of cutoff frequencies $f_T$ between 60 and 100 GHz. All three devices are tested at drain biases of 2 V but the devices that showed the highest $f_T$ at 2 V was also tested at $V_D$ = 2.5 V resulting in $f_T$ = 100 GHz (Some of this data was reported in reference 14).

**Figure 11:** (a) Raman spectrum of graphene on SiC 6H(0001) (same sample corresponding to Table II and Figure 2c,d) after SiC background subtraction. (b) Plot of the 2D peak intensity of the Raman spectrum of the same sample for three different positions on a quarter of a 2-inch wafer (close to the center, intermediate position between center and edge, and close to the edge). For each position, several spectra separated by about 10 μm from each other were taken and plotted.

**Figure 12:** Plot of room-temperature Hall mobility vs. top-gate voltage.